\documentclass[conference]{IEEEtran}
\IEEEoverridecommandlockouts
\usepackage{cite}
\usepackage{amsmath,amssymb,amsfonts}
\usepackage{algorithmic}
\usepackage{graphicx}
\usepackage{textcomp}
\usepackage{xcolor}
\usepackage{float}
\def\BibTeX{{\rm B\kern-.05em{\sc i\kern-.025em b}\kern-.08em
    T\kern-.1667em\lower.7ex\hbox{E}\kern-.125emX}}

\usepackage{fancyhdr}
\begin{document}

\title{AttentioNet: Monitoring Student Attention Type in Learning with EEG-Based Measurement System\\
\thanks{
* These authors contributed equally.\\
\textsuperscript{†} This work was conducted during their affiliation with IIIT-Delhi, India.}
% should not be used} % so delete this line
}
\makeatletter
\newcommand{\linebreakand}{%
  \end{@IEEEauthorhalign}
  \hfill\mbox{}\par
  \mbox{}\hfill\begin{@IEEEauthorhalign}
}
\makeatother

\author{\IEEEauthorblockN{Dhruv Verma\textsuperscript{*†}}
\IEEEauthorblockA{\textit{University of Toronto} \\
Toronto, Canada \\
dhruvverma@cs.toronto.edu}
\and
\IEEEauthorblockN{Sejal Bhalla\textsuperscript{*†}}
\IEEEauthorblockA{\textit{University of Toronto} \\
Toronto, Canada \\
sejal@cs.toronto.edu}
\and
\IEEEauthorblockN{S. V. Sai Santosh\textsuperscript{*†}}
\IEEEauthorblockA{\textit{NVIDIA Corporation} \\
Austin, USA \\
vsiripurapu@nvidia.com}
\linebreakand
\IEEEauthorblockN{Saumya Yadav}
\IEEEauthorblockA{\textit{Human-Machine Interaction Lab}\\
\textit{IIIT-Delhi}\\
Delhi, India \\
saumya@iiitd.ac.in}
\and
\IEEEauthorblockN{Aman Parnami}
\IEEEauthorblockA{\textit{WEAVE Lab}\\
\textit{IIIT-Delhi}\\
Delhi, India \\
aman@iiitd.ac.in}
\and
\IEEEauthorblockN{Jainendra Shukla}
\IEEEauthorblockA{\textit{Human-Machine Interaction Lab}\\
\textit{IIIT-Delhi}\\
Delhi, India \\
jainendra@iiitd.ac.in}
}
\maketitle
\thispagestyle{fancy}

\begin{abstract}
Student attention is an indispensable input for uncovering their goals, intentions, and interests, which prove to be invaluable for a multitude of research areas, ranging from psychology to interactive systems. However, most existing methods to classify attention fail to model its complex nature. To bridge this gap, we propose AttentioNet, a novel Convolutional Neural Network-based approach that utilizes Electroencephalography (EEG) data to classify attention into five states: Selective, Sustained, Divided, Alternating, and relaxed state. We collected a dataset of 20 subjects through standard neuropsychological tasks to elicit different attentional states. The average across-student accuracy of our proposed model at this configuration is 92.3\% (SD=3.04), which is well-suited for end-user applications. Our transfer learning-based approach for personalizing the model to individual subjects effectively addresses the issue of individual variability in EEG signals, resulting in improved performance and adaptability of the model for real-world applications. This represents a significant advancement in the field of EEG-based classification. Experimental results demonstrate that AttentioNet outperforms a popular EEGnet baseline (p-value $<$ 0.05) in both subject-independent and subject-dependent settings, confirming the effectiveness of our proposed approach despite the limitations of our dataset. These results highlight the promising potential of AttentioNet for attention classification using EEG data.
\end{abstract}


\section{Introduction}
Education plays a crucial role in the development of individuals and societies, and effective learning is essential for academic success. The field of education is constantly evolving, and with the advent of advanced technologies, there is a growing interest in exploring innovative methods to enhance educational practices and improve learning outcomes. Student engagement is a crucial factor that influences the quality of learning experiences and outcomes. Engaged students are more likely to be actively involved in their learning, motivated to learn, and able to retain and apply knowledge effectively. Hence, furnishing teachers with feedback on student engagement holds the utmost significance in the realm of education.\par 
Engagement is a multifaceted construct that encompasses three essential components: behavioral engagement, cognitive engagement, and emotional engagement \cite{b2,b3,b4}. Emotional engagement pertains to the affective reactions exhibited by students, such as expressions of happiness or sadness, while behavioral engagement encompasses outward actions, such as asking questions and showing interest during class, as noted in previous research \cite{b5}. On the other hand, cognitive engagement involves internal neuropsychological processes, including attention and investment in learning \cite{b6}. The current study investigates the cognitive engagement of students, which refers to the extent of mental investment in learning, including traits such as reflection, persistence, and experiencing a state of flow. Some indications imply that there may be a positive association between cognitive engagement and academic performance in students \cite{b7,b8}.\par

Attention is a crucial component of cognitive engagement in students, playing a fundamental role in their ability to focus, process information, and actively participate in learning activities. In this study, attention is considered a latent variable of engagement, and its types are determined using neuroanatomic theories, factor analysis of psychometric tests, cognitive processing, and clinical-based models \cite{b9}. The clinical-based model encompasses five distinct types of attention, each with its unique characteristics and relevance to the cognitive processes of students. It emphasizes the multi-dimensional nature of attention, making it highly relevant and appropriate for fine-grained attention modeling. Understanding these different types of attention can be beneficial in shedding light on how students engage and interact with their learning environment:
\begin{itemize}
\item \textbf{Alternating attention:} It reflects mental flexibility in shifting the focus of attention. It allows students to shift their focus between different tasks or stimuli, helping them to manage multiple information sources and switch between different cognitive demands effectively. This can be particularly helpful in complex learning situations that require shifting between different activities or subjects.
\item \textbf{Divided attention:} It represents the ability to respond to multiple tasks simultaneously. It enables students to allocate their cognitive resources to multiple tasks simultaneously. This can be beneficial in situations where students need to multitask or manage multiple information streams, such as when studying with digital resources or taking notes during a lecture.
\item \textbf{Focused attention:} It describes attention to specific tactile, auditory, or visual stimuli. It allows students to concentrate their mental resources on a specific task or stimulus. This type of attention is crucial for deep learning and critical thinking, as it enables students to thoroughly process information, analyze it, and engage in higher-order cognitive processes.
\item \textbf{Selective attention:} It depicts the ability to maintain cognitive focus amidst distracting or competing stimuli. This type of attention enables students to filter out irrelevant distractions and focus on relevant information selectively. This can help students to prioritize important information, ignore distractions, and maintain focus on the task at hand, leading to more efficient and effective learning.
\item \textbf{Sustained attention:} It represents consistent attention during repetitive activities. It allows students to maintain focus and concentration over an extended period of time. This is essential for tasks that require prolonged mental effort, such as reading, problem-solving, or studying. Sustained attention helps students to stay engaged and attentive throughout the learning process, leading to better comprehension and retention of information.
\end{itemize} 
Overt measures of attention in the classroom, such as tracking device usage or gaze patterns, may only capture external stimuli that grab attention, such as a ringing smartphone. However, these techniques do not consider the attentional state influenced by internal stimuli, such as thoughts. In contrast, physiological signals such as Electroencephalography (EEG), functional Near Infrared Spectroscopy (fNIRS), and Electrodermal Activity (EDA) take into account the unfiltered, involuntary changes in the nervous system, thus accounting for the covert state of mind as well [12]. Among these methods, EEG-based approaches have been extensively studied for detecting dynamic mental states due to their high temporal resolution, lack of clinical risk, affordability, and portability.\par
To address the challenges, we propose a novel technique leveraging multi-channel dynamics and temporal dependencies of EEG data from neuropsychological tests to comprehensively model student attention. We demonstrate its efficacy in distinguishing between different types of attention based on the Clinical Model.We conducted a user study (N=20) to evaluate the performance of our approach in detecting the nature of student engagement in accordance with the clinical model. Our study focused on a specific context and population, and we acknowledge that the findings may not be generalizable to all situations. Since EEG data features long-range temporal dependencies, we designed a Convolutional Neural Network (CNN) with self-attention modules, termed as AttentioNet, to classify the different types of attention. CNN was selected for its capability to learn hierarchical representations from complex EEG data, capture spatial patterns, and efficiently learn from limited datasets. To summarize, the key contributions of this paper are as follows:
\begin{itemize}
\item Collection of EEG data from 20 subjects and 5 classes through standard neuropsychological tasks to elicit different attentional states.
\item We introduce AttentioNet, a novel CNN based approach for classifying various attention types using EEG data. Inclusion of multi-channel information in the EEG data, capturing temporal and spatial dependencies for more accurate attention classification.
\item Personalization of the model to individual subjects using a transfer learning-based approach, effectively addressing the issue of individual variability in EEG signals.  
\end{itemize}

\section{Related work}
Previous research has predominantly centered on the detection of engagement using EEG 
\cite{b16, b15, kosmyna2019attentivu}.  While recent works have applied various algorithms on EEG data to achieve an accuracy of 89.4\% in binary classification between attentive and non-attentive/relaxation states \cite{b1} and a maximum of 80.84\% in 3-class classification \cite{b20}. Engagement, which is a multifaceted concept, plays a crucial role in learning and academic achievement. It's essential to recognize that engagement and attention are distinct cognitive processes, although they are often mistakenly considered the same. In the study \cite{b13,b14,b15}, attention and engagement are used interchangeably as indicators of engagement. Also, in the past, there have been efforts to identify any single attention type in application-specific scenarios \cite{b10,b11,b19}. However, to our knowledge, just one other work \cite{b12} has attempted to classify attention according to the clinical model. It uses thermal imaging and remote eye tracking to infer attentional states but suffers from a lack of reliability since both modalities give overt measures of attention. \par

The conventional approach to detecting attention using EEG often involves relying on self-report or external annotator as the ground truth, where participants or external annotators are asked to report their own level of attention subjectively. However, these method has several drawbacks. Firstly, self-report can be influenced by various factors, such as social desirability bias or subjective interpretations, which may result in inaccurate or biased measurements of attention. Secondly, self-report can be influenced by subjective biases and may not always accurately reflect the true attentional state of the individual. Similarly, external annotators may introduce errors or biases in their assessments, leading to less reliable results. \par
 These limitations highlight the need for alternative methods that can provide more objective and reliable measures of attention using EEG. To overcome this in the proposed work, standardized neuropsychological assessments are employed to assess attention, as they comprehensively measure various aspects of attention. These assessments are typically administered by trained psychologists or clinical psychologists and involve task-based evaluations designed to measure attentional processes.

 \begin{figure*}[!ht]
  \centering
  \includegraphics[width=0.8\linewidth]{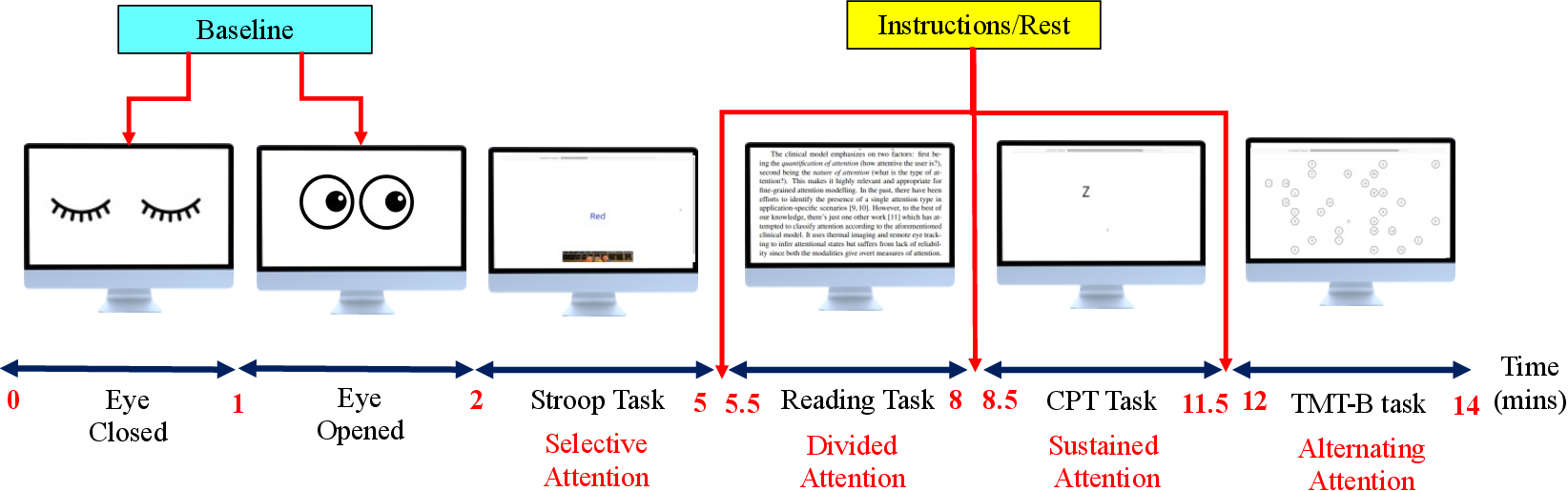}
  \caption{Experiment timeline for AttentioNet data collection.}
  \label{fig1}
\end{figure*}

\section{Study Design}
Training data was collected with ethics approval from the Institutional Review Board of Indraprastha Institute of Information Technology, Delhi, India (IIITD/IRB/9/10/2019-1). Participants underwent an experiment, as outlined in Fig.~\ref{fig1} of the paper. Volunteers did not receive any compensation for participating, and all of them provided informed consent before proceeding with the experiment.

\subsection{Participants}
A total of 20 healthy participants (10 females, 10 males)
within the age range of 18-21 years (mean = 19.58, SD =
0.86) took part in the experiment. Every participant in the
study was an undergraduate student. They all self-reported not
having color blindness, a history of neurological disease, or
hearing disorder, and they had normal or corrected-to-normal
vision.\par

\subsection{Data collection}
In this research study, each participant completed a computerized experiment that involved following specific instructions. It took approximately 14 minutes for each participant to complete the test. The experiment comprised four established neuropsychological tasks that stimulate distinct types of attention, including Selective, Divided, Sustained, and Alternating attention. We omitted focused attention from the study as our primary interest lies in examining attentional patterns over extended periods \cite{b12}. EEG data and facial video recordings of the participants were recorded during the experiment. Additionally, self-reported measures of distraction (applicable to divided attention tasks only) were collected using a Likert Scale. To minimize any potential disruptions to the experimental flow, these ratings were obtained at the end of the experiment by simultaneously presenting facial and screen recordings of the trial to trigger recall. A total of 20 videos were collected, which were used to construct the AttentioNet dataset.\par

Fig.~\ref{fig1} summarizes the AttentioNet experiment. The AttentioNet experiment began with a 1-minute baseline recording with closed eyes, followed by a 1-minute baseline with open eyes for relaxation. The 30-second resting periods and 1-minute intervals for eye-closed and eye-opened states were chosen to ensure reliable data collection and consistent results. Subsequently, the subject performed the neuropsychological tasks, each separated by a 30-second resting period.

\begin{itemize}
\item \textbf{Stroop Test:} The Stroop test is a well-established tool for evaluating selective attention \cite{b21}. It gauges the time taken by individuals to verbally identify the font color of a word that presents an incongruent color. In our study, we utilized a computerized version of the test where participants reported the font color using arrow keys. In the experiment, a slide deck containing the words 'red', 'green', 'blue', and 'yellow' in random order was presented to the subjects. The participants were then instructed to enter the font color of the word displayed using the corresponding keys 'r', 'g', 'b', or 'y'.

\item \textbf{Reading Test:} The experiment was a recreation of the study conducted by in \cite{b10} to assess the state of divided attention. Divided attention was induced by creating an environment where readers could be distracted by external stimuli, in this case, distracting sounds. A total of five reading sessions, each lasting approximately 30 seconds, were conducted. Afterward, participants were asked to rate their level of distraction during each session using a Likert scale ranging from 1-9, with 9 indicating the highest level of distraction. It should be noted that the purpose of these self-reported ratings is to assess the participants' perception of their own distraction levels, rather than serving as the primary measure of attention in the overall study.

\item \textbf{Continous Performance Test(CPT):} The current study utilized the Conners' CPT II test \cite {b22}, which is a task that involves a "No-Go" CPT paradigm. It is a cognitive task that measures sustained attention and divided attention. In this task, participants are presented with a series of stimuli (letters) on a computer screen and are instructed to respond. In the experiment, participants were instructed to press the space bar whenever they see any letter other than "X", indicating a "No-Go" response for "X". The test is typically administered for a duration of approximately 3 minutes. 
\item \textbf{Trail Making Test-B(TMT-B):} TMT, a widely used neuropsychological assessment tool, evaluates cognitive performance and measures various types of attention \cite{b23}. TMT comprises two types: TMT-A, involving number sequencing from 1 to 15, and TMT-B, which requires set-shifting, where subjects alternate between numerical and alphabetic sequences (1-A-2-B-3...) \cite{b24}. For our study, we exclusively employed the TMT-B task to analyze participants' alternating attention. Participants were instructed to select a sequence of alphanumeric bubbles that turned green upon selection. In this assessment, participants are tasked with connecting 40 circled numbers and letters (20 of each) that are randomly distributed, following alternating numerical and alphabetical sequences, akin to 1-A-2-B-3-C... and so on, providing a challenging task for cognitive evaluation.
\end{itemize} 

\section{Methodology for AttentioNet}

\subsection{EEG Data Acquisition and Preprocessing}
For our EEG recording, we utilized the Emotiv EPOC+\footnote{\textbf{\texttt{http://www.emotiv.com/epoc}}} in combination with Emotiv PRO software. The EPOC+ system enabled us to acquire the raw EEG signals wireless via Bluetooth at a high sampling rate of 128 Hz from 14 electrodes strategically positioned on the scalp, including {\texttt{AF3, F7, F3, FC5, T7, P7, O1, O2, P8, T8, FC6, F4, F8,}} and {\texttt{AF4}}. To ensure data quality, the collected raw EEG data were bandpass filtered within the frequency range of 0.2 Hz to 45 Hz. Subsequently, the filtered data was segmented and labeled according to the different attention tasks performed by the subjects. To facilitate further analysis, we epoched the segmented data into frames of 1 second, with an overlap of 750 milliseconds, to capture potential transient changes in the EEG signals during the tasks. This meticulous approach to data pre-processing allows for a more accurate and reliable analysis of the EEG data in our study.

\begin{figure*}[!ht]
    \centering
    \includegraphics[width=1.0\linewidth]{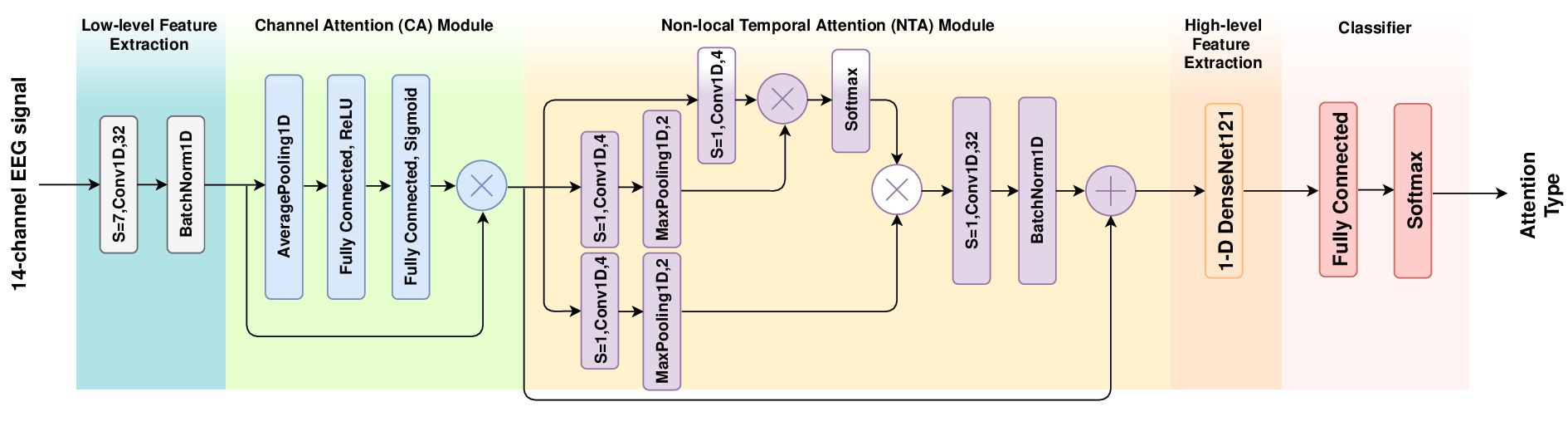}
    \caption{Overview of the AttentioNet Architecture: A Visual Representation of the Attention-based Neural Network Model}
    \label{fig: architecture}
\end{figure*}

\subsection{AttentioNet Architecture}
In this section, we describe the architecture of our proposed model, termed as AttentioNet (see Fig. \ref{fig: architecture}). Given a multi-channel frame of raw EEG signals (time steps, channels) which is called an epoch, the goal is to recognize the attention type (from relaxed, selective, sustained, alternating, and divided attention states) represented by this epoch. Recently, \cite{b29} demonstrated the use of self-attention in robust subject-independent EEG analysis. Motivated by the same, we design our network to encode the temporal and multi-channel dynamics in EEG using an attention mechanism. Our approach draws inspiration from \cite{b30}, which proposes to use this technique for efficient affect recognition using EDA. The architecture consists of 4 blocks which are described as follows:

\subsubsection{{Low-level Feature Extraction}}
In our methodology, we begin with a low-level feature extraction (LFE) module, which plays a crucial role in capturing the fundamental characteristics of the acquired EEG data. We extract low-level feature maps $X_{LFE}$ from the multi-channel input $X$ using a 1-D convolutional layer followed by batch normalization. The 1-D convolutional layer operates on the multi-channel input $X$, which represents a multi-dimensional array with multiple channels. The convolutional layer applies filters to each channel independently, extracting features from each channel separately. This allows the model to capture unique patterns and interactions specific to each channel, while also considering the collective information across channels.
Following the convolutional layer, batch normalization is applied to normalize the output. Batch normalization is a technique that helps stabilize and accelerate the training process of deep neural networks. It normalizes the output across channels, ensuring that features extracted from different channels are on similar scales. The combination of the 1-D convolutional layer and batch normalization allows the proposed network to account for multi-channel dynamics and capture multivariate interactions.

Next, we incorporate a signal attention mechanism on the $X_{LFE}$ to extract temporal and channel-wise features. This attention mechanism enables the model to focus on relevant regions of the EEG signals, capturing important patterns and variations that may be indicative of different cognitive states or tasks. By adaptively attending to different parts of the EEG data, the model can effectively highlight informative features for further analysis.

In the next section, we detail the architecture, including the LFE module and the signal attention mechanism, and how they collaborate to extract meaningful features from the EEG data. This comprehensive approach leverages both low-level and high-level information, resulting in a more robust and discriminative representation for subsequent analysis and interpretation.
\subsubsection{Attention Module}
After extracting low-level features, the feature map is processed using a series of sub-modules, including a channel attention module (CA) and a non-local temporal attention module (NTA).

\textbf{{Channel Attention Module}}
It is widely acknowledged that distinct cognitive and affective states elicit specific responses in different regions of the brain, as evidenced by the EEG electrode activity within those regions. Therefore, investigating the inter-channel (region) relationships and identifying the channels that play a pivotal role in discriminating attention types are critical considerations. \par
The EEG data collected during the experiments comprises 14 channels, each of which may contribute differently to the final prediction. The channel-wise attention mechanism, proven to be portable for various tasks while maintaining excellent performance and robustness to noisy inputs, is utilized to extract the inter-channel relationships. We implement a modified version of the channel attention module inspired by the Squeeze-and-Excitation Block, as proposed in reference \cite{b28}, to suit the unique characteristics of EEG signals. This adaptation includes customizing 1D convolutions designed explicitly for EEG data in order to achieve our desired outcome.\par

We start by performing average temporal pooling on the low-level feature map $X_{LFE}$ to extract channel-wise statistics. These statistics are then passed through a multi-layer perceptron (MLP) with 64 hidden units for further processing. ReLU activation is employed to introduce non-linearity, and Sigmoid activation is applied at the end to generate normalized channel weights ranging from 0 to 1. Finally, the original low-level feature map $X_{LFE}$ is re-weighted by element-wise multiplication with the channel attention weights, resulting in the channel attention-enhanced feature map denoted as $X_{CA}$. Finally, the output of the $X_{CA}$ module can be shown as follows:
\begin{equation}
X_{CA} = Sigmoid(MLP(AvgPool(X_{LFE}))) \cdot X_{LFE}
\end{equation}

\textbf {Non-local Temporal Attention Module (NTA)}
The acquired EEG data in our experiments is sequential, which implies that long-range samples may impact the final outcome differently at different positions. Therefore, it is crucial to investigate these long-range dependencies in the signal to gain a comprehensive understanding of the data. \par
In this context, non-local operations have been proven effective in various applications such as video classification, object detection, instance segmentation, and keypoint detection \cite{b25}. The generalizability and success of non-local operations in these domains have inspired us to apply them to leverage the long-range relationships in neuropsychological signals for attention-type recognition. \par
A generic non-local operation in a deep neural network, as shown in \cite{b25,b26}. The equation defining $\hat{x_i}$ is as follows:
\begin{equation}
\hat{x_i} = \frac{1}{c(x)} \sum_{\forall_j} f(x_i, x_j) g(x_j),
\end{equation}

In this equation, $x$ represents the input feature map, denoted as $X_{CA}$ in our case. The index $i$ represents the target position in time for the output, while the index $j$ refers to the set of all possible positions that contribute to the response $x_i$. The function $f(\cdot)$ extracts the relationship between index $i$ and all possible positions $j$, while $g(\cdot)$ re-weights the input features. The function $c(\cdot)$ represents a normalization function applied before obtaining the final output.

In this work, we use the convolution operation coupled with a max-pooling operation for the linear operation $g(\cdot)$ as:
\begin{equation}
g(x_j) = {MaxPool}(W_gx_j),
\end{equation}

where $W_g$ represents the convolution kernel. For the pairwise operation $f(\cdot)$, we apply the commonly used Embedded Gaussian formulation \cite{b27} (a widely used measure for computing similarity in an embedding space) to define $f(\cdot)$ as:
\begin{equation}
f(x_i, x_j) = e^{\theta x_i^T \phi x_j}
\end{equation}

where $\theta(\cdot)$ and $\phi(\cdot)$ represent convolution operations. For the normalization operator $c(\cdot)$, we adopt the adaptation from reference \cite{b25}, where it is set as $\sum_{\forall_j} f(x_i, x_j)$, and the final pairwise operation is performed using the softmax operation $\frac{1}{c(x)} f(x_i, x_j)$. The overall operation of the NTA block can be expressed as:
\begin{multline}
        X_{NL} = Softmax(X_{CA}^{T}W_{\theta}^{T}MaxPool(W_{\phi}X_{CA}))\\
        MaxPool(W_{g}X_{CA})
\end{multline}

where $W_\theta$ and $W_\phi$ are convolution kernels and finally, the output of the NTA module can be shown as: 
\begin{equation}
X_{NTA} = W_wF_{NL} + F_{CA}
\end{equation}
Where $W_w$ denotes the 1D convolution operation followed by batch normalization. The output $X_{NTA}$ refers to the re-weighted result obtained from the Attention Module.
% $X_{NTA}$ is the re-weighted output of the Attention Module.

\subsubsection{{High-level Feature Extraction}}
We employ DenseNet-121 \cite{b31} as the backbone for extracting high-level feature maps, as it offers improved gradient flow and tractable optimization. It is worth noting that other state-of-the-art feature extraction backbones such as VGG, ResNet, or GoogLeNet exist, but a comparative analysis of these networks is not the primary focus of this work. To make DenseNet-121\footnote{A detailed diagram for the modified DenseNet-121 is skipped due to space constraints.} compatible with the input signal features, we modify its original architecture by replacing the 2D convolutions with 1D convolutions. The backbone takes the input $X_{NTA}$ and produces a feature vector denoted as $X_{HFE}$.

\subsubsection{Classifier}
Finally, the feature vector $X_{HFE}$ is fed into a Fully Connected Layer with Softmax activation, consisting of 5 units, for the purpose of classifying attention into their respective types.

\subsection{Model Training}
After pre-processing the data (as described in Section 4.A), we obtained a total of $62,530$ windows, each with a length of 128 samples, across $20$ subjects and five classes (as detailed in Section 4.B). To train our neural network, a leave-one-subject-out scheme was adopted, where the data of one subject was reserved for validation while the remaining subjects' data was utilized for training. The Adam optimizer \cite{b32} was employed, with a learning rate of 0.001, and was decayed by a factor of 10 if the validation loss plateaued after an epoch. The Categorical Cross-entropy function served as the loss function. Mini-batches of size 32 were used for training, with model checkpoints saved at the epoch with the highest validation accuracy.

\section{Experiment Results}
To assess the performance of our approach in detecting student attention type, we conducted experiments in two different settings: subject-independent and subject-dependent (adaptive) evaluations. Additionally, since this is the first attempt to categorize attention using EEG data, we compare our model's performance to EEGNet \cite{b33}.

EEGNet is a architecture specifically designed for analyzing EEG data. It addresses the unique challenges posed by EEG signals, such as non-stationarity and temporal dependencies, by incorporating depthwise and separable convolutions. These convolutions efficiently capture spatial and temporal patterns in the EEG data. Moreover, EEGNet employs a parallel filter-bank structure to process different frequency bands separately, enabling the model to extract multi-scale information present in EEG signals effectively.

Due to its effectiveness in handling EEG data, EEGNet is widely used in EEG-based classification tasks, including motor imagery classification \cite{hernandez2021motor}, emotion recognition \cite{tunnell2022novel}, and various cognitive tasks \cite{b33}. Researchers often use EEGNet as a benchmark model to evaluate and compare the performance of their proposed approaches in EEG data analysis.

\subsection{Subject Independent Evaluation} 
To evaluate the model's performance without subject-specific data, we employed a leave-one-subject-out (LOSO) cross-validation approach. The dataset was split based on individual subjects, where data from 19 participants were used for training, and the data of one participant was held out for validation. This process was repeated for each subject, ensuring a comprehensive assessment of the model's performance across different subjects.

The average LOSO accuracy for AttentioNet in the 5-class attention classification task was 53.75\% (SD=4.58, chance=20\%), while for EEGNet, it was 43.8\% (SD=6.20). Our approach significantly outperformed EEGNet, and we confirmed this with a Students' T-test ($p:   1.5 \times 10^{-6}$).

However, we recognized that considering the inherent variability of EEG signals across participants could lead to further improvements in performance. As a result, we decided to adopt a subject-dependent (adaptive) approach, enabling us to personalize the model for each individual. This adaptability takes into account the unique characteristics of each subject's EEG data, which has been demonstrated to lead to significant enhancements in the model's performance \cite{nath2020comparitive}.

\subsection{Subject Dependent Evaluation} 
To account for individual variability, we employed a transfer learning-based approach using the LOSO cross-validation method. For each left-out test subject, we fine-tuned a base model pre-trained on the data from the remaining subjects. This fine-tuning process involved using a subset of the left-out subject's data. To determine the optimal size of this subset, indicating the number of samples needed for effective personalization, we incrementally added 10 seconds of the subject's data per class. After each step, we fine-tuned the model and evaluated the performance of AttentioNet in comparison to EEGNet. 

\begin{figure}[!t]
    \centering
    \includegraphics[width=0.8\linewidth]{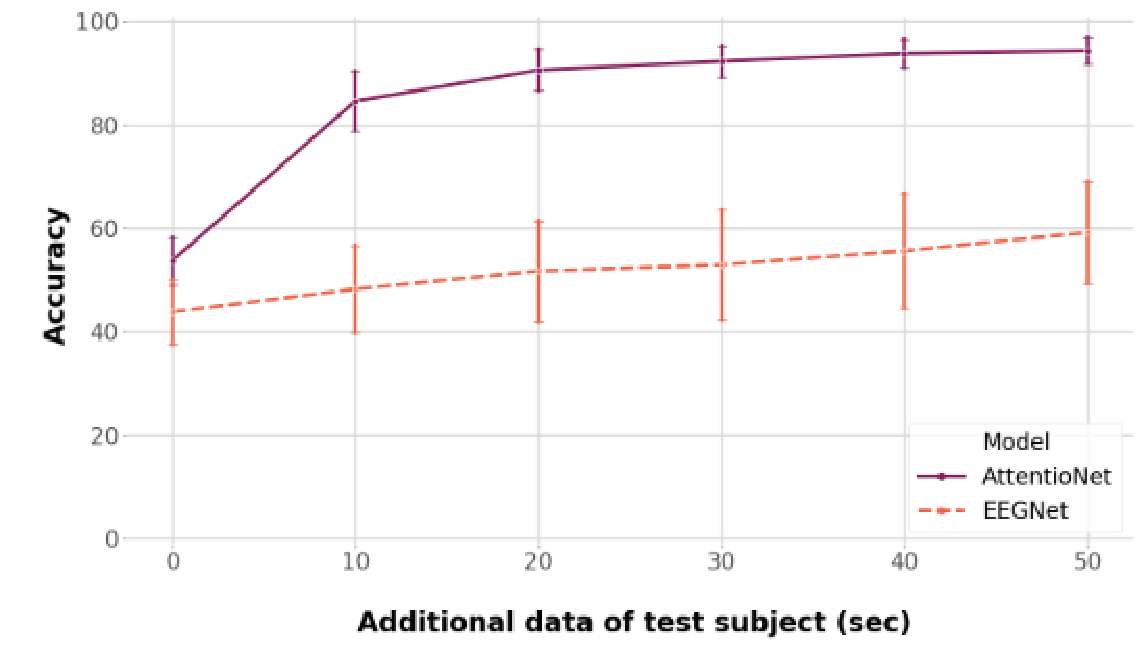}
    \caption{Performance of AttentioNet and EEGNet tuned with different amounts of data from the test subject. Error bar indicates the standard error.}
    \label{fig: samplesresult}
\end{figure}

As depicted in Fig.~\ref{fig: samplesresult}, the transfer learning approach led to a significant improvement in performance for both networks, with AttentioNet exhibiting a notably higher increase. Notably, using 30 seconds of data per class was found to be sufficient to personalize the model for any given subject. At this configuration, the average across-subject accuracy of AttentioNet was 92.3\% (SD=3.04), making it well-suited for end-user applications. Any further increase in the tuning data beyond this point resulted in an insignificant gain in accuracy, suggesting that additional data can be traded off to minimize the data required for fine-tuning the base model.

Fig.~\ref{fig: subjectsresults} summarizes the performance of our approach and EEGNet, each tuned according to the aforementioned optimal configuration, in a 5-class classification task for each subject. This analysis highlights the effectiveness of our subject-dependent adaptive approach in achieving highly accurate attention classification results across various individuals.

\begin{figure}[!h]
    \centering
    \includegraphics[width=0.9\linewidth]{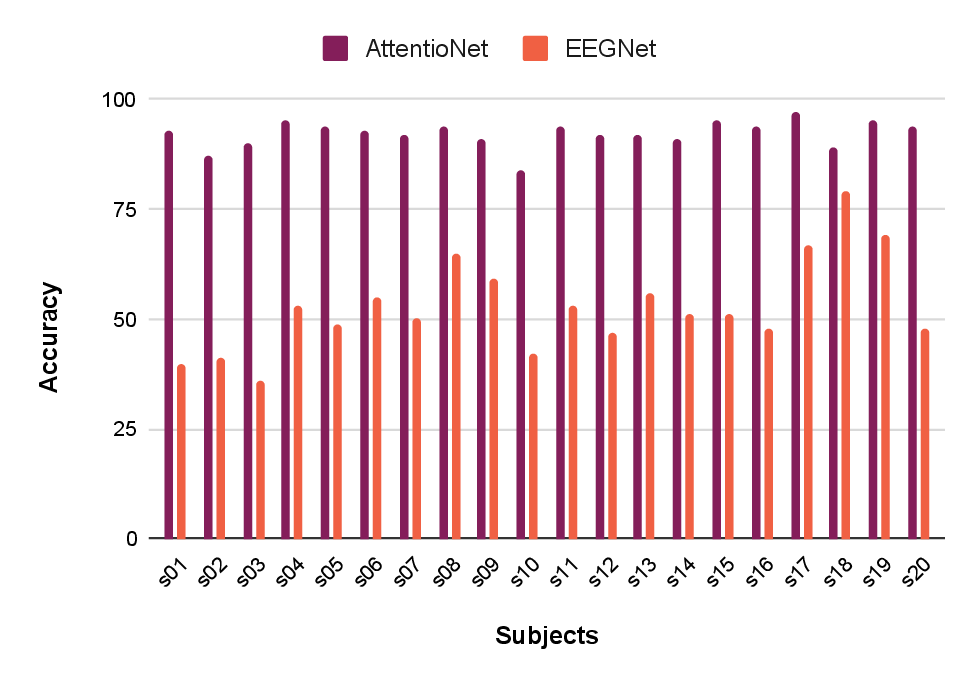}
    \caption{Comparison of EEGNet and AttentioNet, tuned with 30 seconds of subject data per class, for each subject.}
    \label{fig: subjectsresults}
\end{figure}

\section{Conclusion}
In this work, we demonstrate the effectiveness of a novel attention classification mechanism that utilizes deep neural networks to analyze EEG data and differentiate between different attention states, as specified by the clinical model of attention. Unlike previous research that treated attention as a unidimensional variable, our approach acknowledges the multidimensional nature of human attention. To evaluate the performance of our approach, we collected data from 20 students who performed a range of controlled neuropsychological tasks, specifically designed to elicit distinct attention states. For classifying attention types, we introduced AttentioNet, a CNN-based architecture that incorporates non-local temporal and channel-wise attention mechanisms. This integration enables robust classification of attention types using EEG data. Comparing our model's performance to EEGNet, a well-established EEG baseline, we observed that AttentioNet outperformed EEGNet with average accuracies of 53.75\% and 92.3\% in subject-independent and subject-dependent settings, respectively (the latter achieved through personalization with 30 seconds of EEG data). Furthermore, our subject-dependent analysis highlighted the importance of personalization in EEG-based attention monitoring, underscoring its potential for enhancing performance in real-world attention-based applications. 

\section{Limitations and Future Work}
As with any study, our research also comes with certain limitations that warrant discussion and consideration. 

Firstly, our dataset comprises a relatively small number of subjects, primarily consisting of students. To develop a more widely applicable and generalizable system, future work should focus on expanding the dataset to include a more diverse range of subjects.

Secondly, the controlled settings used to evoke attentional states may not fully replicate the complexities of real-world situations. Addressing this limitation involves conducting experiments in more naturalistic settings to capture user responses in authentic, everyday scenarios.

Thirdly, our data collection was confined to a single session. To achieve a more comprehensive evaluation of personalized algorithms, future research should incorporate data from multiple sessions to assess performance more holistically.

Despite these limitations, our findings illustrate the promising potential of EEG-based systems for attention monitoring and classification. Looking ahead, we envision integrating our approach with portable EEG devices like AttentivU glasses \cite{kosmyna2019attentivu}, facilitating real-time attention measurement. Additionally, by incorporating advanced sensing technologies, our work has the potential to provide enhanced insights into teaching, thereby improving learning outcomes in educational settings.

\section{Acknowledgements.} This research was supported by the Centre for Design and New Media (a TCS Foundation Initiative supported by Tata Consultancy Services) and the Infosys Centre for Artificial Intelligence at IIIT-Delhi, India. Gratitude is also extended to all the participants who contributed to this study.

\section{Ethical Impact Statement}
Our research delves into the use of deep neural networks to analyze human cognitive states, particularly attention, with a focus on young populations like students. Consequently, our work extends beyond technological advancements and raises various ethical implications that demand careful consideration. 

Firstly, the collection of EEG data from human subjects for research purposes necessitates strict adherence to ethical guidelines. This includes obtaining informed consent from participants, safeguarding their confidentiality, and ensuring minimal discomfort or harm during data acquisition. To ensure ethical compliance, our study was approved by the IRB at our institution.

Secondly, we recognize the potential impact on privacy and data security. EEG data is sensitive and can reveal personal information about an individual's cognitive processes and attentional states. To safeguard privacy, measures were implemented, such as data anonymization, secure storage, and restricted access to the data.

Thirdly, the limitations of the dataset used for training and evaluation must be acknowledged, as it may not fully represent the diversity of the general population in terms of age, gender, ethnicity, and other demographic factors. Efforts were made to carefully interpret the findings and avoid overgeneralization.

Moreover, the potential for bias in the proposed model has been carefully evaluated. Convolutional Neural Networks (CNNs) are known to learn patterns from data, and if the training data is biased, it can lead to biased predictions. Thus, the model's performance across different demographic groups was thoroughly assessed and ensured that it does not perpetuate existing biases or discriminate against specific populations.

Furthermore, the intended use and application of AttentioNet have been considered. If AttentioNet is used in real-world settings for monitoring and classifying attention states, it may raise concerns about surveillance, autonomy, and fairness. Ethical considerations will be taken into account when deploying AttentioNet in practical applications.

In conclusion, while AttentioNet represents a significant advancement in EEG-based attention classification, it is imperative to adhere to ethical principles throughout the research process. Transparent reporting of limitations, potential biases, and ethical considerations is crucial to ensuring the responsible and ethical use of this technology. By upholding ethical standards, we strive to maximize the positive impact of our research while minimizing potential risks or unintended consequences.

\bibliography{references}
\end{document}